%Paper: hep-th/9510206
%From: avramidi@math-inf.uni-greifswald.d400.de
%Date: Sat, 28 Oct 1995 16:20:50 +0100

------------------------------ Start of body part 2

%
% Paper:  Algorithms for the calculation of the heat kernel coefficients
%
% Proceedings of III Leipzig Workshop on ``Quantum Field Theory
% under the Influence of External Conditions'',
% Leipzig, 18-22 Sept. 1995,
%
% Plain TeX
%
% Authors: 			I. G. Avramidi and R. Schimming
%
% Mailing address:		Department of Mathematics
%		 		University of Greifswald
%				Jahnstr. 15a
%				D-17489 Greifswald
%				Germany
%
% E-mail:			avramidi@math-inf.uni-greifswald.d400.de
% Tel:				(03834)-864614
% FAX:				(03834)-864615; 861372
%

\magnification=1200
\tolerance=2000
\hbadness=2000
\overfullrule=0pt

% fonts

\font\gross=cmbx10 scaled \magstep2
\font\mittel=cmbx10 scaled\magstep1
\font\sf=cmss10
\font\pl=cmssq8 scaled \magstep1
\font\titlesf=cmssbx10 scaled \magstep2
\font\authorsf=cmss10 scaled \magstep1
\font\sc=cmcsc10

\def\RR{{\rm I\!R}}

\def\a{\alpha}
\def\b{\beta}
\def\g{\gamma}
\def\d{\delta}

\def\l{\lambda}
\def\m{\mu}
\def\n{\nu}

\def\s{\sigma}

\def\na{\nabla}
\def\tr{{\rm tr\,}}

\def\const{{\rm const}}

% ----------------------

\nopagenumbers
\def\rightheadline{\it\title\qquad\hfill\rm\folio}
\def\leftheadline{\rm\folio\hfill\it\qquad\author}
\headline={\ifnum\pageno>1{\ifodd\pageno\rightheadline\else\leftheadline\fi}
\else\fi}

\def\author{I. G. Avramidi and R. Schimming}
\def\title{Algorithms for heat kernel coefficients}

% titlepage

\null
\vskip-1.5cm
\hskip6.2cm{ \hrulefill }
\vskip-.55cm
\hskip6.2cm{ \hrulefill }
%\vskip1.5mm
\smallskip
\hskip6.2cm{{\pl \ University of Greifswald (October, 1995)}}
%\smallskip
\smallskip
%\vskip0.1mm
\hskip6.2cm{ \hrulefill }
\vskip-.55cm
\hskip6.2cm{ \hrulefill }
\bigskip
\bigskip
\hskip6.2cm{\ to appear in: }
\bigskip
\hskip6.2cm{\sf \ Proceedings of the IIIrd Workshop}
\smallskip
\hskip6.2cm{\sf \  ``Quantum Field Theory under}

\hskip6.2cm{\sf \  Influence of External Conditions''}
\smallskip
\hskip6.2cm{\sf \ Leipzig, September 18-22, 1995}

\vskip3cm
\centerline{\titlesf Algorithms for the calculation of the
}\centerline{\titlesf
heat kernel coefficients}

\bigskip
\bigskip

\centerline{{\authorsf I.\ G.\ Avramidi}
\footnote{*}{Alexander von Humboldt Fellow. On leave of absence from Research
Institute for Physics, Rostov State University,  Stachki 194, 344104
Rostov-on-Don, Russia. {\sf E-mail: avramidi@math-inf.uni-greifswald.d400.de}}
and
{{\authorsf R.\ Schimming}}
\footnote{\$}{\sf E-mail: schimming@math-inf.uni-greifswald.d400.de}
}
\medskip
\centerline{\it Institut f\"ur Mathematik/Informatik,
E.-M.-Arndt-Universit\"at,}
\centerline{\it 17487 Greifswald, Germany}
\bigskip

\vskip3cm
{\narrower
\noindent
We present a brief overview of several approaches for calculating the local
asymptotic expansion of the heat kernel for Laplace-type operators. The
different methods developed in the papers of both authors some time ago are
described in more detail.
\par}
\par

\vfill\eject
\pageno=1

\centerline{\gross Algorithms for the calculation of the
}\centerline{\gross
heat kernel coefficients}
\bigskip

\centerline{${\bf I.\ G.\ Avramidi}$
\footnote{*}{Alexander von Humboldt Fellow. On leave of absence from Research
Institute for Physics, Rostov State University,  Stachki 194, 344104
Rostov-on-Don, Russia. {\sf E-mail: avramidi@math-inf.uni-greifswald.d400.de}}
and
$\underline{{\bf R.\ Schimming}}$
\footnote{\$}{\sf E-mail: schimming@math-inf.uni-greifswald.d400.de}
}
\medskip
\centerline{\it Institut f\"ur Mathematik/Informatik,
E.-M.-Arndt-Universit\"at,}
\centerline{\it 17487 Greifswald, Germany}
\bigskip

\leftline{\mittel 1. Laplace-type differential operators}
\medskip
The convential Laplacian $\Delta=div\,grad$ on the Euclidean space $\RR^n$
admits a natural generalization to a Riemannian manifold $(M,g)$:
$$
\Delta=g^{ab}\na_a\na_b.
$$
Here
$$
g=g_{ab}dx^adx^b, \qquad (g^{ab}):=(g_{ab})^{-1}
$$
with respect to local coordinates $x^a=x^1,x^2,\dots,x^d$, and  $\na$ is the
Levi-Civita derivative to $g$ acting on scalar or tensor fields. Another
far-reaching generalization is a {\it Laplace-type} linear differential
operator of the form
$$
L=g^{ab}D_aD_b+W
\eqno(1)
$$
acting on the smooth sections of a vector bundle $E$ over the manifold $M$. A
covariant derivative $D$ of these sections and a section $W$ of the
endomorphism bundle $End \, E$ to $E$ enter the expression for $L$.

An intrinsic definition, alternative to the coordinate-dependent definition
(1), can be given
[54].
Let the map $ad_{f}$ for a smooth scalar field $f\in C^{\infty}(M)$ act on any
linear operator $L$ according to
$$
ad_{f} L:=[L,f]\equiv Lf-f L.
$$
(Here and in the following a scalar field is identified with the operator of
multiplication by this scalar field.) A linear operator $L$ acting on
$\varphi\in C^{\infty}(E)$ is called a Laplace-type differential operator if
$$
(ad_{f})^2L=2|df|^2.
$$
It can be shown then that
$$
2<df,D\varphi>:=(ad_f L)\varphi-(\Delta f)\varphi
$$
defines a covariant derivative $D$ on $C^\infty(E)$
and that
$$
W\varphi:=L\varphi-g^{ab}D_aD_b\varphi
$$
defines an endomorphism $W\in C^\infty(End\,E)$. We use the notation
$$
<v,w>:=g^{ab}v_aw_b, \qquad |v|^2:=<v,v>
$$
for bundle-valued one-forms $v=v_adx^a$, $w=w_adx^a$.

Clearly, the class of Laplace-type operators $L$ is form-invariant under
diffeomorphisms $\Phi$, when the objects $g$, $D$, $W$ are carried along with
$\Phi$.
This class is also form-invariant under {\it gauge transformations}
$$
\bar L:=\Lambda^{-1}L\Lambda,
$$
where $\Lambda\in C^\infty(Aut\,E)$ is a section of the automorphism bundle to
$E$, and under {\it conformal transformations}
$$
\bar L:=e^{-(m+2)f}Le^{mf},
$$
where $f\in C^\infty(M)$ and $m:=(d-2)/2$, $d$ being the dimension of $M$.

We have
$$
\bar g=g, \qquad \bar D=\Lambda^{-1}D\Lambda, \qquad
\bar W=\Lambda^{-1}W\Lambda
$$
under a gauge transformation with $\Lambda$, and
$$
\bar g=e^{2f}g, \qquad \bar
W=e^{-2f}\left(W+I\cdot e^{-mf}\Delta e^{mf}\right)
$$
under a conformal transformation with $f$. Multiple covariant derivatives $D$
of sections of $E$ behave like multiple Levi-Civita derivatives of a scalar
field under conformal transformations. In particular,
$$
\bar D\varphi=D\varphi \ \ {\rm for} \ \varphi\in C^\infty(E).
$$
For details cf.
[54,33].

\bigskip
{\leftline{\mittel 2. The heat kernel coefficients}
\medskip

Some sequence of locally defined two-point quantities $H_k=H_k(x,x')$
($k=0,1,2,\dots$) is associated to any Laplace-type operator $L$.
\smallskip
\noindent
{\sl Proposition 1.} A formal series expression
$$
K=K(t;x,x')=(4\pi t)^{-d/2}\exp\left(-{\sigma(x,x')\over 2t}\right)
\sum\limits_{k=0}^\infty H_k(x,x')t^k
\eqno(2)
$$
solves the heat equation
$$
{\partial K\over \partial t}=L K
$$
with the initial condition
$$
K(+0;x,x')=\d(x,x')
$$
if and only if
$$
\left({\cal D}+\m\right)H_0=0, \qquad H_0(x,x)=I,
\eqno(3)
$$
and
$$
\left({\cal D}+k+\m\right)H_k=L H_{k-1}, \qquad {\rm for}\  k=1,2,\dots.
\eqno(4)
$$
Here $\sigma(x,x')$ is the so called geodetic interval defined as one half the
square of the distance along the geodesic between the points $x$ and $x'$,
${\cal D}$ is a first-order linear differential operator given by
$$
{\cal D}\varphi=<d\sigma,D\varphi>,
$$
$I$ is the identity endomorphism, and $\m={1\over 2}(\Delta\sigma-d)$.

All the analysis in this paper is purely local. This means that we fix some
small regular region of the manifold $M$ and consider the points $x$ and $x'$
to lie inside this region.
\smallskip
\noindent
{\sl Proposition 2.} There is a neighbourhood $N$ of the diagonal of $M\times
M$ and a unique smooth system of solutions $H_k=H_k(x,x')$ in $N$ of the
differential recursion system (4). These $H_k=H_k(x,x')$ are called the {\it
Hadamard coefficients} to $L$.

Note that the quantities $H_k=H_k(x,x')$ are also widely known under the name
Hadamard-\-Minakshisundaram-\-De Witt-\-Seeley (HMDS or HAMIDEW coefficients),
according to the papers of these authors
[36,42,43,21,22,51,52],
or heat kernel coefficients,  the last being used in the title of this paper.
Hadamard
[36]
introduced them already in 1923 for scalar operators $L$ and established the
essential properties.

\bigskip
\leftline{\mittel 3. Methods for effective calculations}
\medskip

We want to classify the methods for an effective calculation of the Taylor
coefficients $[D^nH_k]$ ($p,k=0,1,2,\dots$) of the Hadamard coefficients or,
which means a little more, of the diagonal values of their multiple covariant
derivatives with respect to $x$, $[D_{a_1}D_{a_2}\cdots D_{a_n}H_k]$
($n,k=0,1,2,\dots$). The square brackets [\ ] applied to a two-point quantity
means the restriction to the diagonal of $M\times M$.
Actually, we arrange the methods to five groups.

\smallskip
{\sc I. Taylor expansion in normal coordinates.}
Naturally, this is a method which comes first to the mind. It has been used
with great skill by G\"unther, W\"unsch, McLenaghan and others
[31-33,65,41].

\smallskip
{\sc II. Manifestly covariant methods.}
The Taylor expansion in normal coordinates can be reinterpreted in a
coordinate-independent manner. So, if one works with covariant derivatives
instead of partial derivatives it should be possible to replace the method I.
by manifestly covariant formulas. In fact, Synge
[60,61]
and De Witt and Brehme
[21,22]
made the first steps in this direction. Their work can be continued in two ways
to an algorithm for the Taylor coefficients $[D^nH_k]$, which are given by the
diagonal values of the {\it symmetrized} covariant derivatives of $H_k$, or to
an algorithm for the diagonal values of the ({\it non-symmetrized}) covariant
derivatives. We will present such algorithms, developed by the authors, in the
next sections.

{\sc III. Invariant-theoretical methods.}
Let $R_{abcd}$ denote the components of the Riemannian curvature of $(M,g)$ and
$F_{ab}$ be the components of the curvature of the bundle connection $D$,
$$
[D_a,D_b]\varphi=F_{ab}\varphi \qquad {\rm for}\ \varphi\in C^\infty(E).
$$
The components of $[D^nH_k]$ or of the diagonal values of the multiple
covariant derivatives of the $H_k$, $[D_{a_1}D_{a_2}\cdots D_{a_n}H_k]$,
exhibit a controlled behavior under diffeomorphisms and gauge transformations.
This implies that they are just polynomials in $g_{ab}$, $g^{ab}$, $R_{abcd}$,
$F_{ab}$, $\na_{a_1}R_{abcd}$, $D_{a_1}F_{ab}$, $\na_{a_1}\na_{a_2}R_{abcd}$,
$D_{a_1}D_{a_2}F_{ab}$, etc. Some careful analysis shows that the coefficients
of these polynomials are determined by the functorial properties under several
decompositions: Riemannian product $(M,g)$=$(M_1,g_1)\times(M_2,g_2)$, direct
vector bundle sum $E=E_1\oplus E_2$, tensor product of vector bundles
$E=E_1\otimes E_2$, and splitting of the potential term $W=W_1+W_2$.
Gilkey
[26-29]
extensively applied the invariant-theoretical method as we call this analysis.
Let us mention also the papers
[65,33,25].

\smallskip
{\sc IV. Heat semigroup methods.}
If $L$ is elliptic, i.e. the metric is positive definite, and if the manifold
$M$ is compact, then the linear operator $\exp(tL)$ is well defined and forms a
semigroup, the so-called heat semigroup. (Let us mention that there is also a
construction of the heat semigroup for non-compact $M$ under certain additional
conditions.)
%Anyway for perturbational purposes it can be considered as a formal
The linear operator $\exp(tL)$ has a kernel $K=K(t;x,x')$. The latter admits
the formal series (2) as an asymptotic series as $t\to+0$, cf.
[42,43,18,28].

In quantum field theory there have been developed several perturbational
techniques for handling a semigroup of the form $\exp(tL)$. Let us mention the
Feynman path integrals, graph methods, the Dyson formula etc. Such methods can
produce results for the kernel $K=K(t;x,x')$ of $\exp(tL)$ and, hence, for the
Hadamard coefficients. Osborn, Zuk, Nepomechie and others used this approach
[44-48,69-72,23].
In recent papers of one of the authors (I.G.A.) the whole heat kernel diagonal
$K(t;x,x)$ (not only the $H_k(x,x)$) in low-energy approximation was
constructed using the semigroup operator approach
[8-14].

The semigroup $\exp(tL)$ has also a probabilistic interpretation: it describes
some stochastic process. In particular, $\exp(tL)$ on scalar fields describes
Brownian motion; the general $\exp(tL)$ belongs to some more complicated Markov
process. The probabilistic approach gives further insights; let us only mention
the Feynman-Kac formula and let us quote the papers
[17,39,19,20].

\smallskip
{\sc V. Pseudodifferential operator methods.} Hadamard in his book
[36]
constructed an asymptotic expansion for the Green function to $L$; he called it
elementary solution. Let $L$ be elliptic and replace it by $L-\l I$, where the
parameter $\l$ belongs to the resolvent set of $L$, i.e. $\l$ is not a spectral
value of $L$. Hadamard's construction for $L-\l I$ can be reinterpreted in
terms of pseudodifferential operators as a parametrix construction. The
coefficients of the asymptotic series are just the Hadamard coefficients, up to
numerical factors. Knowing this, one can produce results for the $H_k$
($k=0,1,2,\dots$) by means of refined pseudodifferential operator methods.
Gilkey, Fulling and Kennedy, Gusynin, Obukhov and others did this succesfully
[27,28,24,35,46,16,64].

Note that our systematics considers only {\it general} methods.
Additionally to them, there are {\it special} methods for restricted classes of
Riemannian manifolds, vector bundles and/or differential operators. Let us only
mention the harmonic analysis on Lie groups or homogeneous manifolds: a
spectral analysis of group-invariant differential operators by means of
representation theory.

\bigskip
\leftline{\mittel 4. An algorithm for the Taylor coefficients of the $H_k$}
\medskip
\noindent
The exposition in this section is due to the papers of one of the authors
(I.G.A.)
[2,3,6,7].
It is not difficult to show that
$$
H_0(x,x')=\Delta^{1/2}(x,x'){\cal P}(x,x').
$$
where
$\Delta(x,x')$ is the Van-Vleck-Morette determinant and ${\cal P}(x,x')$ is the
parallel displacement operator with respect to $D$ along the geodesic.
\noindent
The modified coefficients
$$
a_k:={(-1)^k\over k!}H_0^{-1}H_k
$$
fulfill the simplified differential recursion system
[22]
$$
a_0=I, \qquad \left(1+{1\over k}{\cal D}\right)a_k=Ma_{k-1}
\eqno(5)
$$
where
$$
M:=-H_0^{-1}LH_0.
\eqno(6)
$$

We assume that there is a unique geodesic connecting the arguments $x$ and $x'$
of our two-point functions and consider all two-point quantities to be
single-valued analytic functions.

\par\noindent
The formal solution of the recursion relations (5) has the form
$$
        a_k= \left(1+{1\over k}{\cal D}\right)^{-1}M
        \left(1+{1\over k-1}
        {\cal D}\right)^{-1}M\cdots(1+{\cal D})^{-1}M
$$
Let us expand $a_k$ in a covariant Taylor series
$$
a_k=\sum_{n\ge 0}\vert n><n\vert a_k>,
$$
where some compact notation is used. Namely, let $grad'\sigma$ denote the
vector field with components
$$
\sigma^{\mu'}=g^{\m'\n'}\sigma_{\n'}\equiv g^{\m'\n'}\na_{\n'}\sigma,
$$
$$
(grad'\s)^n:=grad'\s\otimes\cdots\otimes grad'\s
$$
be the $n$-fold (symmetric) tensor product and
$$
\vert n>:={(-1)^n\over n!}(grad'\s)^n.
$$

Let further the linear functionals $<n|$ \ (dual to the system of functions
$|n>$) be de\-f\-ined by
$$
<n|\varphi>:=[D^n\varphi]
$$
for a two-point field $\varphi$, where $D^n=D D\cdots D$ is the multiple
symmetric covariant derivative with respect to $x$, and the square brackets [\
] denote, as before, the restriction to the diagonal.
Then $|n><n|\varphi>$ denotes the inner product
$$
|n><n|\varphi>=(|n>)^{\m_1\dots\m_n}(<n|\varphi>)_{\m_1\dots\m_n}.
$$

\noindent
{}From the remarkable eigenvalue formula
$$
{\cal D}\vert n > =n\vert n>
$$
one can obtain an expression for the Taylor coefficients of the $a_k$, namely
$$
        <n\vert a_k>=\sum_{n_1,\cdots,n_{k-1}\ge 0}\left(1+{{n}\over {k}}
        \right)^{-1}\left(1+{{n_{k-1}}\over {k-1}}\right)^{-1}
        \cdots\left(1+n_1\right)^{-1}
$$
$$
        \times <n\vert M\vert n_{k-1}><n_{k-1}\vert M\vert n_{k-2}>\cdots
        <n_1\vert M\vert 0>,
$$
where $<m\vert M\vert n>$  are the `matrix elements' of the operator
$M$ (6)
$$
<m\vert M\vert n>
=\left[D^m M{{(-1)^n}\over {n!}}
(grad'\s)^n\right].
\eqno(7)
$$
\noindent
The matrix elements $<m\vert M\vert n>$ are tensors with components
$M_{\ \ \ \ \ \ \ \m_1\dots\m_m}^{\n_1\dots\n_n}$ which are symmetric both in
upper and lower indices. The inner product of the matrix elements is again such
a tensor with components
$$
(<n|M|k><k|M|m>)^{\n_1\dots\n_n}_{\ \ \ \ \ \ \m_1\dots\m_m}
=M_{\ \ \ \ \ \ \ \l_1\dots\l_k}^{\n_1\dots\n_n}
M_{\ \ \ \ \ \ \ \m_1\dots\m_m}^{\l_1\dots\l_k}.
$$
These tensors are expressible in terms of diagonal values of derivatives of
basic two-point quantities. The components of these matrix elements were found
in [2,3,6,7].

In order to present explicit formulas for these matrix elements we use a
calculus of matrix-valued, vector-valued and $End E$-valued symmetric
differential forms. So we introduce first matrix-valued symmetric forms
$K_{(n)}=(K^\n_{\ \m (n)})$:
$$
K^\n_{\ \ \m(n)}=\nabla_{(\mu_1}\cdots\nabla_{\mu_{n-2}}
R^\n_{ \ \mu_{n-1}\vert\m\vert\mu_n)}dx^{\m_1}\vee\cdots\vee dx^{\m_n},
$$
where $\vee $ denotes the symmetric tensor product of symmetric forms.
Then we define the matrix-valued symmetric forms
$\g_{(n)}=(\g^\a_{\ \b(n)})$, $\eta_{(n)}=(\eta^\a_{\ \b(n)})$ and
$(X^{\m\n}_{\ \ (n)})$ by
$$
\g_{(n)}
         =\sum_{1\le k \le [{n\over 2}] }(-1)^{k+1}
       \sum_{{n_1,\cdots ,n_k\ge2 \atop n_1+\cdots +n_k=n}}
       N^{-1}(n_1,\cdots,n_k)
       K_{(n_k)}\vee\cdots\vee K_{(n_2)}\vee K_{(n_1)},
$$
where
$$
\eqalignno{
&N(n_1,\cdots ,n_k)={(n+1)\over (n-1)!}(n_1-2)!\cdots (n_k-2)!&\cr
&\times n_1(n_1+1)(n_1+n_2)(n_1+n_2+1)\cdots (n_1+\cdots +n_{k-1})
(n_1+\cdots +n_{k-1}+1),&\cr}
$$
and
$$
\eta_{(n)} =-\sum_{1\le k\le [{n\over 2}]}
        \sum_{{n_1,   \cdots ,n_k\ge 2\atop n_1+\cdots +n_k=n}}
        {n!\over n_1!\cdots    n_k!}\ \
        \g_{(n_k)}\vee\cdots\vee\g_{(n_2)}\vee\g_{(n_1)},
$$
$$
X^{\m\n}_{\ \ (n)}=\sum_{0\le k\le n}{n\choose k}\eta^{(\m}_{\ \
\a(k)}\vee\eta^{\n)\a}_{\ \ \ (n-k)},
$$
and scalar symmetric forms $\zeta_{(n)}$
$$
        \zeta_{(n)} =
        \sum_{1\le k\le [{n\over 2}]}
        {1\over 2k}
        \sum_{n_1,\cdots ,n_k\ge 2\atop n_1+\cdots +n_k=n}
        {n!\over n_1!\cdots n_k!}
        \tr\,\left(\g_{(n_k)}\vee\cdots\vee\g_{(n_2)}\vee\g_{(n_1)}\right).
$$
Next, let us introduce the vector-valued (and $End E$-valued) symmetric forms
$F_{(n)}=(F_{\m(n)})$:
$$
F_{\m(n)}=D_{(\mu_1}\cdots D_{\mu_{n-1}}{F}_{|\m|\mu_n)}dx^{\m_1}\vee\cdots\vee
dx^{\m_n}
$$
and $A_{(n)}=(A_{\m(n)})$
$$
A_{(n)}={{n}\over {n+1}}\left\{F_{(n)}
-\sum_{1\le k\le n-2}{n-1\choose k-1}{F}_{(k)}\vee\gamma_{(n-k)}\right\},
$$
and $End E$-valued symmetric forms $W_{(n)}$
$$
W_{(n)}=W_{\mu_1\cdots\mu_n}dx^{\m_1}\vee\cdots\vee dx^{\m_n}
=D_{(\mu_1}\cdots D_{\mu_n)}Wdx^{\m_1}\vee\cdots\vee dx^{\m_n}.
$$
Using the introduced quantities we define finally some more symmetric forms
$$
X^{\n\a}_{\ \ \a(n)}=X^{\n\a}_{\ \ \a\m_1\dots\m_n}
dx^{\m_1}\vee\cdots\vee dx^{\m_n},
$$
and
$$
A_{\a\b(n)}=A_{\a\b\m_1\dots\m_n}
dx^{\m_1}\vee\cdots\vee dx^{\m_n}, \qquad
\zeta_{\a\b(n)}=\zeta_{\a\b\m_1\dots\m_n}
dx^{\m_1}\vee\cdots\vee dx^{\m_n}.
$$
where $X^{\m\n}_{\ \ \m_1\dots\m_n}$, $A_{\m\ \m_1\dots\m_n}$ and
$\zeta_{\m_1\dots\m_n}$ are the components of the forms $X^{\m\n}_{\ \ (n)}$,
$A_{\m(n)}$ and $\zeta_{(n)}$.
The components of the matrix elements $<m\vert M\vert n>$ are given then by
$$
        <m\vert M\vert n>=0 \qquad\qquad {\rm for} \quad n>m+2\quad
           {\rm and} \quad   n=m+1,
$$
$$
\eqalignno{
M_{\ \ \ \ \ \ \ \m_1\dots\m_m}^{\n_1\dots\n_n}=&
        {m\choose n}\delta^{\nu_1\cdots\nu_n}_{(\mu_1\cdots\mu_n}
        Z_{\m_{n+1}\cdots\mu_m)}
+{m\choose n-1}\delta^{(\nu_1\cdots\nu_{n-1}}_{(\mu_1\cdots\mu_{n-1}}
        Y^{\nu_n)}_{\ \ \ \mu_n\cdots \mu_m)} &\cr
&
-{m\choose n-2}
	I\,\delta^{(\nu_1\cdots\nu_{n-2}}_{(\mu_1\cdots\mu_{n-2}}
        X^{\nu_{n-1}\nu_n)}_{\ \ \ \ \ \ \ \ \mu_{n-1}\cdots\mu_m)},
&\cr}
$$
where $X^{\m\n}_{\ \ \mu_1\cdots\mu_n}$, $Y^\nu_{\ \ \mu_1\cdots\mu_n}$ and
$Z_{\mu_1\cdots\mu_n}$ are the components of the forms $X^{\mu\nu}_{\ \ \
(n)}$, $Y^\nu_{\ \ (n)}$ and $Z_{(n)}$:
$$
\eqalignno{
Y^\nu_{\ \ (n)} &=
- I\,X^{\nu\alpha}_{\ \ \ \a(n)}
+2\sum_{0\le k\le n}{n\choose k}
X^{\n\m}_{\ \ (k)}\vee A_{\m (n-k)},  & \cr
Z_{(n)} &=
-W_{(n)}
&\cr&
+\sum_{0\le k\le n}{n\choose k}\biggl\{
X^{\alpha\beta}_{\ \ (k)}\vee\left(
-A_{\a\b(n-k)}
+I\,\zeta_{\a\b(n-k)}\right)
%&\cr&
+X^{\b\a}_{\ \ \ \alpha(k)}\vee
\left(- A_{\b (m)}
+I\,\zeta_{\b (m)}\right)
\biggr\} & \cr
&\
+\sum_{m,k\ge 0\atop m+k\le n}{{n!}\over {k!m!(n-k-m)!}}
X^{\alpha\beta}_{\ \ (k)}\vee\bigl(A_{\a (m)}\vee A_{\b (n-k-m)}
-I\,\zeta_{\a(m)}\vee\zeta_{\b(n-k-m)}\bigr).&\cr}
$$

\bigskip
{\leftline{\mittel 5. An algorithm for the derivatives of the $H_k$.}
\medskip

The following is due to the papers of the other author (R.Sch.)
[53-58].
Let us begin with a new tensor notation: indices at a tensor symbol shall not
denote the components but the valences, that means the entries of the tensor
taken as a multilinear functional. Positive integers $1,2,3\dots$ are preferred
indices for the valences. Thus, $u_{12\dots p}$ denotes a covariant $p$-tensor,
$u^{12\dots p}$ denotes a contravariant $p$-tensor, $u_{12\dots p}+ v_{12\dots
p}$ is the sum of two tensors, $u_{12\dots p}v_{p+1,\dots, p+q}$ is a tensor
product, the natural action of a permutation $\pi$ of $1,2,\dots,p$ on a
$p-$tensor is
$$
\pi u_{12\dots p}=u_{\pi(1)\pi(2)\dots \pi(p)}.
$$
More generally, if $\Pi$ is a subset of the symmetric group $S_p$ (of
permutations of $1,2,\dots,p$), then
$$
\Pi u_{12\dots p}:=\sum\limits_{\pi\in \Pi}\pi u_{12\dots p}.
$$

We are interested here in special permutations: a $q$-shuffle of $1,2,\dots,p$
is a $\pi\in S_p$ such that
$$
\pi(1)<\pi(2)<\cdots<\pi(q)
\qquad {\rm and} \qquad
\pi(q+1)<\pi(q+2)<\cdots<\pi(p).
$$
The action of the set of all $q$-shuffles on a covariant tensor $u_{12\dots p}$
is denoted by
$$
u_{\underline{12\dots q}\ \underline{q+1\dots p}}.
$$
For example:
$$
u_{\underline{1}\ \underline{23}}=u_{123}+u_{213}+u_{312},
$$
$$
u_{\underline{12}\ \underline{34}}=u_{1234}+u_{1324}+u_{1423}
+u_{2314}+u_{2413}+u_{3412}.
$$

Let us define tensors $S_{a12\dots p}$ ($p\ge 3$) and $End\,E$-valued tensors
$M_{12\dots p}$ ($p\ge 2$) by
$$
S_{a123}:=R_{a123}, \qquad S_{a1234}:= R_{a123;4}
$$
$$
S_{a12\dots p}:=R_{a12\underline{3};\underline{45\cdots p}}
+\sum\limits^{p-3}_{q=2}R_{a1b2;\underline{3\cdots q}}
S^b_{\ \underline{q+1\dots p}},
$$
$$
M_{12}:=F_{12}, \qquad M_{123}:=F_{1\underline{2};\underline{3}},
$$
$$
M_{12\dots p}:=F_{1\underline{2};\underline{34\dots p}}
+\sum\limits^{p-3}_{q=1}F_{a1;\underline{23\dots q}}
S^a_{\ \underline{q+1\dots p}},
$$
where indices after a semicolon express covariant derivatives.
We define also
$$
s_{a12\dots p}:=[\s_{;a12\dots p}]+S_{a12\dots p}
$$
and
$$
m_{12\dots p}:=[\m_{;12\dots p}]-M_{12\dots p}.
$$

The diagonal values of the derivatives of the Hadamard coefficients are then
recursively given by
$$
(p+k)[H_{k;12\dots p}]=[(L\,H_{k-1})_{;12\dots p})]
$$
$$
-\sum\limits_{q=2}^pm_{\underline{12\dots p}}[H_{k;\underline{q+1,\dots p}}]
-\sum\limits_{q=3}^ps^a_{\ \underline{12\dots q}}[H_{k;a\underline{q+1\dots
p}}],
$$
where
$$
(LH_k)_{;12\dots p}=H_{k;\ a12\dots p}^{\ \ a}
+\sum\limits_{q=0}^{p}{p\choose q} W_{;\underline{12\dots
q}}H_{k;\underline{q+1 \dots p}}.
$$

\bigskip
\leftline{\bf 6. Discussion}
\vglue0pt
\medskip
\vglue0pt
Let us add some historical remarks on the explicit calculation of the diagonal
values of the Hadamard coefficients and their Taylor coefficients.
Already in the thirties Heisenberg and Euler
[37]
and Mathisson
[40]
have calculated $[H_1]$, $[DH_1]$, $[D^2H_1]$ for Laplace-type operators in
flat space,
$L=g^{ab}D_aD_b+W$, $g^{ab}=\const$.
For scalar operators in curved space let us mention H\"older
[38]
for $[H_1]$, G\"unther
[31]
for $[H_1]$, $[H_2]$, Sakai
[50]
for $[H_3]$, Amsterdamski, Berkin and O'Connor
[1]
for $[H_4]$, G\"unther
[32]
for $[DH_1]$ and $[D^2H_1]$ and W\"unsch
[66-68]
for $[DH_2]$ and $[D^2H_2]$. For non-scalar operators
in flat space we mention
[62]
for $[H_4]$ and
[63]
for $[H_5]$.

Results for the general Laplace-type operator have been obtained by De Witt
[22]:
$[H_1]$, $[H_2]$, Gilkey
[26]:
$[H_1]$,$[H_2]$,$[H_3]$, one of the authors (I.G.A.)
[2,3,6,7]:
$[H_1]$,$[H_2]$,$[H_3]$,$[H_4]$, the other author (R.Sch.)
[53,54]:
$[DH_1]$,$[D^2H_1]$,$[D^3H_1]$, W\"unsch
[65]:
$[D^4H_1]$, and others.

The Hadamard coefficients $H_k$ or rather the one-point quantities $[D^nH_k]$
or $[D_{a_1}D_{a_2}\cdots D_{a_n}H_k]$ derived from them have many appearances
or applications:
\item{---} Huygens' principle for hyperbolic $L$
[40,31-33,41,53-55,65-68],
\item{---}  Spectral geometry for elliptic $L$ and compact Riemannian $(M,g)$
with positive definite metric $g$
[42,43,18],
\item{---} ``Heat kernel proofs''  of the index theorems
[34,28],
\item{---}  Zeta-function regularization,
\item{---}  Regularized energy-momentum tensor,
\item{---}  Korteveg-De Vries hierarchy
[56,15].

It should be noted that there are some other problems in mathematical physics
that require a similar technique for investigation of transport equations.
These are:
\item{---} expansion of the metric and other quantities in normal coordinates
[33],
\item{---}  harmonic manifolds
[49] and harmonic differential operators
[58],
\item{---}  volume problems in the sense of
[30] (to read geometric information from the volume of geodesic balls,
truncated light cones, ...)
[30,57],
\item{---} Brownian motion.

\bigskip
\leftline{\bf Acknowledgements}
\bigskip
The work of I.G.A. was supported by the Alexander von Humboldt Foundation.
He is grateful to J. Eichhorn for his hospitality at the University of
Greifswald.

\bigskip
\leftline{\bf References}
\vglue0pt
\medskip
\vglue0pt

\item{[1]} P. Amsterdamski, A. L. Berkin and D. J. O'Connor, Class. Quant.
           Grav. {\bf 6} (1989) 1981

\item{[2]} I. G. Avramidi, {\it The covariant methods for calculation of the
           effective action in quantum field theory and the investigation of
           higher derivative quantum gravity}, PhD Thesis, Moscow State
           University, Moscow, 1986; hep-th/9510140
\item{[3]} I. G. Avramidi, Teor. Mat. Fiz. {\bf 79} (1989) 219; (Theor. Math.
Phys. {\bf 79} (1989) 494)
\item{[4]} I. G. Avramidi, Yad. Fiz. {\bf 49} (1989) 1185; (Sov. J. Nucl.
Phys., {\bf 49} (1989) 735)
\item{[5]} I. G. Avramidi, Phys. Lett. B {\bf 236} (1990) 443
\item{[6]} I. G. Avramidi, Phys. Lett. B {\bf 238} (1990) 92
\item{[7]} I. G. Avramidi, Nucl. Phys. B {\bf 355} (1991) 712
\item{[8]} I. G. Avramidi, Phys. Lett. B {\bf 305} (1993) 27
\item{[9]} I. G. Avramidi, Phys. Lett. B {\bf 336} (1994) 171
\item{[10]} I.~G.~Avramidi, {\it Covariant methods for calculating the
low-energy effective action in quantum field theory and quantum gravity},
University of Greifswald (1994), gr-qc/9403036
\item{[11]} I. G. Avramidi, {\it A new algebraic approach for calculating the
heat kernel in quantum gravity}, University of Greifswald (1994),
hep-th/9406047, J. Math. Phys. {\bf 37} (1) (1996),  to appear
\item{[12]} I. G. Avramidi, {\it  New algebraic methods for calculating the
heat kernel and the effective action in quantum gravity and gauge theories},
gr-qc/9408028, in: {\it `Heat Kernel Techniques and Quantum Gravity'}, {\it
Discourses in Mathematics and Its Applications},  No.~4, Ed.  S.~A. Fulling,
Texas A\&M University, (College Station, Texas, 1995), to appear
\item{[13]} I. G. Avramidi, J. Math. Phys. {\bf 36} (1995) 5055
\item{[14]} I. G. Avramidi, {\it Covariant approximation schemes for
calculation of the heat kernel in quantum field theory}, University of
Greifswald (1995), hep-th/9509075, Proc. Int. Sem. ``Quantum Gravity'', Moscow,
June 12-19, 1995, to appear
\item{[15]} I. G. Avramidi and R. Schimming, J. Math. Phys. {\bf 36} (1995)
5042

\item{[16]} A. O. Barvinsky and G. A. Vilkovisky, Phys. Rep. {\bf 119} (1985) 1

\item{[17]} C. Bellaiche, Asterisque, {\bf 84/85} (1981) 151

\item{[18]} M. P. Berger, P. Gauduchon and E. Mazet, {\it Le spectre d'une
variete riemannienne}, Lecture Notes in Math. {\bf 194}, Berlin 1971

\item{[19]} J. M Bismut, J. Funct. Anal. {\bf 57} (1984) 56

\item{[20]} J. M. Bismut, Comm. Math. Phys. {\bf 98} (1985) 213

\item{[21]} B. S. De Witt and R. W. Brehme, Ann. Phys. {\bf 9} (1960) 220
\item{[22]} B. S. De Witt, {\it Dynamical theory of groups and fields},
(Gordon and Breach , N.Y., 1965)

\item{[23]} Y. Fujiwara, T. A. Osborn and S. F. Wilk, Phys. Rev. A {\bf 25}
(1982) 14

\item{[24]} S. Fulling and G. Kennedy, Transac. Am. Math. Soc. {\bf 310} (1988)
583
\item{[25]} S. Fulling, R. C. King, B. G. Wybourne and C. C. Cummins, Class.
Quant. Grav. {\bf 9} (1992) 1151

\item{[26]} P. B. Gilkey, J. Diff. Geom. {\bf 10} (1975) 601
\item{[27]} P. B. Gilkey, Compositio Math. {\bf 38} (1979) 201
\item{[28]} P. B. Gilkey, {\it Invariance theory , the heat equation and the
Atiyah-Singer index theorem}, (Publish or Perish, Wilmington, 1984)
\item{[29]} P. B. Gilkey, Contemp. Math. {\bf 73} (1988) 79

\item{[30]} A. Gray and L. Vanhecke, Acta Math. {\bf 142} (1979) 157

\item{[31]} P. G\"unther, Ber. Verhand. S\"achs. Akad. d. Wiss. Leipzig {\bf
100} (1952) Heft 2
\item{[32]} P. G\"unther, Math. Nachr. {\bf 22} (1960) 285
\item{[33]} P. G\"unther, {\it Huygens' Principle and Hyperbolic Equations},
(Academic Press, San Diego, 1988)
\item{[34]} P. G\"unther and R. Schimming, J. Diff. Geom. {\bf 12} (1977) 599

\item{[35]} V. P. Gusynin, Nucl Phys. B {\bf 333} (1990) 296

\item{[36]} J. Hadamard, {\it Lectures on Cauchy's problem}, (Yale Univ. Press,
New Haven, 1923)

\item{[37]} W. Heisenberg and H. Euler, Z. Phys. {\bf 98} (1936) 714
\item{[38]} E. H\"older, Ber. Verh. S\"achs. Akad. Wiss. Leipzig, {\bf 99}
(1938) 55

\item{[39]} T. Jacobson, J. Math. Phys. {\bf 26} (1985) 1600

\item{[40]}  M. Matthisson, Acta Math. {\bf 71} (1939) 249
\item{[41]} R. G. McLenaghan, Proc. Camb. Phil. Soc., {\bf 65} (1969) 139
\item{[42]} S. Minakshisundaram and A. Plejel, Canad. J. Math. {\bf 1} (1949)
242
\item{[43]} S. Minakshisundaram, J. Indian Math. Soc. {\bf 17} (1953) 158
\item{[44]} F. H. Molzahn and T. A. Osborn, J. Math. Phys. {\bf 27} (1986) 88
\item{[45]} R. I. Nepomechie, Phys. Rev. D {\bf 31} (1985) 3291
\item{[46]} Yu. N. Obukhov, Nucl. Phys. B {\bf 212} (1983) 237
\item{[47]} T. A. Osborn and F. H. Molzahn, Phys. Rev. A {\bf 34} (1986) 1696
\item{[48]} T. A. Osborn and R. A. Corns, J. Math. Phys. {\bf 26} (1985) 453
\item{[49]} H. S. Ruse, A. G. Walker, T. J. Willmore,
		{\it Harmonic spaces}, (Edizioni Cremonese, Roma, 1961)
\item{[50]} T. Sakai, Tohoku Math. J. {\bf 23} (1971) 589
\item{[51]} R. T. Seeley, AMS, Proc. Symp. Pure Math. 10 (1967) 288
\item{[52]} R. T. Seeley, Am. J. Math. {\bf 91} (1969) 889

\item{[53]} R. Schimming, Ukrainsk. Mat. Z., {\bf 29} (1977) 351
\item{[54]} R. Schimming, Beitr\"age zur Analysis, {\bf 11} (1978) 45
\item{[55]} R. Schimming, Beitr\"age zur Analysis, {\bf 15} (1981) 77
\item{[56]} R. Schimming, Z. f. Analysis u. ihre Anwend., {\bf 7} (1988) 263
\item{[57]} R. Schimming, Archivum Math. Brno, {\bf 24} (1988) 5
\item{[58]} R. Schimming, Forum Math., {\bf 3} (1991) 177
\item{[59]} R. Schimming, {\it Calculation of the heat kernel coefficients},
in: {\it Analysis, Geometry and Groups. A Riemann Legacy Volume}, Eds. H. M.
Srivastava and Th. M. Rassias, (Hadronic Press, Palm Harbour, 1993), p. 627

\item{[60]} J. L. Synge, Proc. London Math. Soc. {\bf 32} (1931) 241
\item{[61]} J. L. Synge, {\it Relativity. The general theory}, (North Holland,
Amsterdam, 1960)

\item{[62]} A. I. Vainstein, V. I. Zakharov, V. A. Novikov and M. A.Shifman,
           Yad. Fiz.  39 (1984) 124
\item{[63]} A. E. M. Van de Ven, Nucl. Phys. B250 (1985) 593
\item{[64]} H. Widom, Bull. Sci. Math. 104 (1980) 19

\item{[65]}  V. W\"unsch, Math. Nachr. {\bf 47} (1970) 131
\item{[66]}  V. W\"unsch, Math. Nachr. {\bf 73} (1976) 37
\item{[67]}  V. W\"unsch, Beitr\"age zur Analysis {\bf 13} (1979) 147
\item{[68]}  V. W\"unsch, Math. Nachr. {\bf 120} (1985) 175

\item{[69]} J. A. Zuk, Phys. Rev. D {\bf 32} (1985) 2650
\item{[70]} J. A. Zuk, Phys. Rev. D {\bf 33} (1986) 3645
\item{[71]} J. A. Zuk, Phys. Rev. D {\bf 34} (1986) 1791
\item{[72]} J. A. Zuk, Nucl. Phys. B {\bf 280} (1987) 125

\bye